\documentclass{mathincs}
\usepackage{bbm}        
\usepackage{bm}         
\usepackage{multirow}

 \theoremstyle{definition}
 
 \theoremstyle{remark}

 \numberwithin{equation}{section}

\def\D#1{D_{\!#1}}
\def\S#1{S_{\!#1}}
\def\d{\partial}
\def\K{\mathbb K}

\def\O{\mathbb O}

\def\rmd{\,\mathrm{d}}  
\def\w{{\bm w}}
\def\dd{\bm{\partial}}
\def\aalpha{\bm{\alpha}}
\def\bbeta{\bm{\beta}}
\def\ggamma{\bm{\gamma}}

\begin{document}
%
%
%
\title{A Fast Approach to Creative Telescoping} 

\author{Christoph Koutschan}

\address{%
Department of Mathematics\\
Tulane University\\
New Orleans, LA 70118}

\email{Koutschan@risc.uni-linz.ac.at}

\thanks{supported by NFS-DMS 0070567 as a postdoctoral fellow, and by the Austrian Science Fund (FWF): P20162-N18\\
The final publication is available at www.springerlink.com, DOI: 10.1007/s11786-010-0055-0.}

\subjclass{Primary 68W30; Secondary 33F10}

\keywords{holonomic functions, special functions, symbolic integration, symbolic summation,
creative telescoping, Ore algebra, WZ theory}

\date{January 8, 2010}


\begin{abstract}
In this note we reinvestigate the task of computing creative
telescoping relations in differential-difference operator algebras.
Our approach is based on an ansatz that explicitly includes the
denominators of the delta parts.  We contribute several ideas of how to
make an implementation of this approach reasonably fast and provide
such an implementation.  A selection of examples shows that it can be
superior to existing methods by a large factor.
\end{abstract}

\maketitle

\section{Introduction}

The method of creative telescoping nowadays is one of the central tools in computer
algebra for attacking definite integration and summation problems. Zeilberger with his
celebrated holonomic systems approach~\cite{Zeilberger90} was the
first to recognize its potential for making these tasks algorithmic
for a large class of functions.
In the realm of holonomic functions, several algorithms for computing
creative telescoping relations have been developed in the past. 
The methodology described here is not an algorithm in the strict sense
because it involves some heuristics. But since it works pretty well on nontrivial 
examples we found it worth to be written down. Additionally we believe that 
it is the method of choice for really big examples. Our implementation
is contained in the Mathematica package \texttt{HolonomicFunctions}
as the command \texttt{FindCreativeTelescoping}. The package can be downloaded
from the RISC combinatorics software webpage:
\begin{center}
\fbox{http://www.risc.uni-linz.ac.at/research/combinat/software/}
\end{center}

Throughout this paper we will work in the following setting.  We
assume that a function~$f$ to be integrated or summed satisfies some
linear difference-differential relations which we represent in a
suitable operator algebra (Ore algebra).  We use the symbol~$\D{x}$ to
denote the derivation operator w.r.t.~$x$ and~$\S{n}$ for the shift
operator w.r.t.~$n$.  Such an algebra can be viewed as a polynomial
ring in the respective operators, with coefficients being rational
functions in the corresponding variables, subject to the commutation
rules $\D{x}x=x\D{x}+1$ and $\S{n}n=n\S{n}+\S{n}$.  Ideally, all the
relations for~$f$ generate a $\d$-finite left ideal, i.e., a
zero-dimensional left ideal in the operator algebra.  If
additionally~$f$ is holonomic (a notion that can be made formal by
$D$-module theory), then the existence of creative telescoping
relations is guaranteed by theory (i.e., by the elimination property
of holonomic modules).  Chyzak, Kauers, and
Salvy~\cite{ChyzakKauersSalvy09} have shown that creative telescoping
is also possible for higher-dimensional ideals under certain
conditions.  We tacitly assume that any input to a creative
telescoping algorithm is $\d$-finite and holonomic, and that it is
given as a left Gr\"obner basis~$G$ of the annihilating ideal of~$f$.

The main concern of this paper is to compute creative telescoping relations
for an integrand (resp. summand) $f(v,\w)$ where~$v$ is the integration (resp. summation) variable
and~$\w=w_1,w_2,\dots$ are some additional parameters. In other words, we are looking for
annihilating operators for $f(v,\w)$ of the form 
\begin{equation}\label{ct}
  P(\w,\dd_{\w}) + \tilde{\d}_v\cdot Q(v,\d_v,\w,\dd_{\w})
\end{equation}
where~$\d_v$ and~$\tilde{\d}_v$ stand for operators acting on the variable~$v$ 
($\d_v=\tilde{\d}_v=\D{v}$ in the case of an integral, and 
$\d_v=\S{v}$ and $\tilde{\d}_v=\S{v}-1$ in the summation case), 
and~$\dd_{\w}$ are the operators $\d_{w_1},\d_{w_2},\dots$ that correspond to the extra parameters.
We will refer to~$P$ as the principal part (also known as the telescoper), and to~$Q$ as the delta part.
From~\eqref{ct} it is immediate to derive relations for the definite integral (resp. sum),
which in general can be inhomogeneous.
Similarly, multiple integrals and sums can be done by creative telescoping
relations that correspond to an operator of the form
\begin{equation}\label{ctmult}
  P(\w,\dd_{\w}) + \tilde{\d}_{v_1}\!\cdot Q_1(\bm{v},\w,\dd_{\bm{v}},\dd_{\w}) + 
                   \tilde{\d}_{v_2}\!\cdot Q_2(\bm{v},\w,\dd_{\bm{v}},\dd_{\w}) + \dots
\end{equation}
where $\bm{v}=v_1,v_2,\dots$ denote the integration (resp. summation) variables
(also mixed cases are possible); we will use the notation 
$\bm{v}^{\aalpha}=v_1^{\alpha_1}v_2^{\alpha_2}\cdots$.
Note that in general the application of~\eqref{ctmult} yields a relation
with inhomogeneous right-hand side that consists of integrals (resp. sums) 
of one dimension lower. These can be treated recursively in the same way.

\section{Description of the method}

The known algorithms for computing creative telescoping relations for
holonomic functions are either based on elimination (e.g., by rewrite
rules / Gr\"obner bases) or on the use of an ansatz with undetermined
coefficients. Zeilberger's slow algorithm~\cite{Zeilberger90} 
and Takayama's algorithm~\cite{Takayama90a,Takayama90b,ChyzakSalvy98}
fall into the first category.  Their advantage is that they can deal
with multiple integrals and sums, but elimination can be a very
difficult task and moreover, it is not guaranteed that they deliver
the smallest relations that exist in the given annihilating ideal. A
relatively small (and hypergeometric!) example that so far has
resisted the elimination approach is given in
Section~\ref{secAndrewsPaule}. Since the algorithms that we are going
to discuss here fall into the second category, we do not want to go
into further detail with elimination-based algorithms.

All the other algorithms make an ansatz with undetermined
coefficients. Reduction modulo the Gr\"obner basis~$G$ gives its
normal form representation. For the creative telescoping relation to
be in the ideal generated by~$G$, its normal form must be identically
zero. Hence equating all coefficients of the normal form to zero gives
rise to a system of equations that can be solved for the undetermined
coefficients. The algorithms described below differ only in the shape
of the ansatz.

A classical algorithm that is based on an ansatz has been proposed by
Chyzak~\cite{Chyzak00}. It can only be applied to single integrals or
single sums, and has to be used in an iterative way for multiple
ones. With the notation of~\eqref{ct} its ansatz is of the following form:
\begin{equation}\label{chyzakAnsatz}
    \sum_{\bbeta\in B}p_{\bbeta}(\w)\dd_{\w}^{\bbeta} 
  \quad+\quad \tilde{\d}_v\cdot
    \sum_{\ggamma\in U}q_{\ggamma}(v,\w)(\d_v,\dd_{\w})^{\ggamma} 
\end{equation}
where~$B$ is a finite multi-index set and~$U$ is the finite set of
multi-indices that correspond to the monomials under the stairs of the
Gr\"obner basis~$G$.  The unknown $p_{\bbeta}(\w)$ are rational
functions in~$\K(\w)$, and the unknown $q_{\ggamma}(v,\w)$ are
rational functions in~$\K(v,\w)$.  When the
ansatz~\eqref{chyzakAnsatz} is written in standard operator
representation, i.e., when~$\tilde{\d}_v$ is commuted to the right, we
encounter derivatives (resp. shifts) of the $q_{\ggamma}(v,\w)$ with
respect to~$v$.  Finally, we end up with a coupled linear first-order
system of differential (resp. difference) equations.  All
implementations of Chyzak's algorithm that we know of\footnote{Besides
our Mathematica package \texttt{HolonomicFunctions}, this refers to
Chyzak's implementation which is part of the Maple package
\texttt{Mgfun} (http://algo.inria.fr/libraries/).}  uncouple this
system and then solve the resulting scalar equations one by
one. Experience shows that these steps can be extremely costly, in
particular when~$U$ is big. As an alternative to uncoupling there are
algorithms for directly solving such coupled systems, proposed by
Abramov and Barkatou~\cite{AbramovBarkatou98,Barkatou99}.  A
comparison of how these methods perform on big creative telescoping
examples could be an interesting research topic for the future.
However, in this article we want to follow a different way of
bypassing the bottleneck, namely by means of a different ansatz that
does not lead to a coupled system.

In~\cite{Koutschan09,Koutschan10a} a first step into this direction has been
taken by means of a ``polynomial ansatz'' of the form
\begin{equation}\label{polyAnsatz}
    \sum_{\bbeta\in B}p_{\bbeta}(\w)\dd_{\w}^{\bbeta}
  \quad+\quad\tilde{\d}_{v_1}\cdot
    \sum_{\ggamma\in C_1}\sum_{\aalpha\in A_1}q_{1,\aalpha,\ggamma}(\w)\,\bm{v}^{\aalpha}\,(\dd_{\bm{v}},\dd_{\w})^{\ggamma}
  +\quad\dots
\end{equation}
where~$A_i$,~$B$, and~$C_i$ are finite sets of multi-indices, and the
dots hide terms with $\tilde{\d}_{v_2},\dots$.  The
unknown~$p_{\bbeta}$ and~$q_{i,\aalpha,\ggamma}$ to solve for are
rational functions in the surviving variables~$\w$ and they can be
computed using pure linear algebra, without any uncoupling needed
(since commuting the $\tilde{\d}_{v_i}$ to the right will not affect
them).  Note also that the summation variables will not occur in the
denominators.  The price that we pay is that the shape of the ansatz
is not at all clear from the beginning: The sets~$A_i$,~$B$, and~$C_i$
need to be fixed, whereas in Chyzak's algorithm we have to loop only
over the support of the principal part (the set~$B$).  Another
drawback of this ansatz is the fact that the delta parts in this
denominator-free representation usually have not only much bigger
supports ($|C_i|>|U|$), but also higher polynomial degrees and larger
integer coefficients than the reduced representation returned by
Chyzak's algorithm\footnote{To give some quantitative results: in the
TSPP example, see Section~\ref{secConclusion}, we ended up with
polynomial degrees up to~$764$ and integer coefficients with up
to~$378$ digits, whereas the reduced representation has degree~$120$
and at most $74$-digit integers.}.  Of course, once found, the
solution of the polynomial ansatz can be transformed to reduced
representation yielding the same result as Chyzak's algorithm in the
one-dimensional case (just reduce the delta part with the Gr\"obner
basis~$G$). So we have now got rid of the uncoupling problem, but the
above observation suggests that the result of the polynomial ansatz is
blown up unnecessarily and that we can do even better.

And in fact, we can! The only thing we have to do is to include the
denominators of the delta parts into the ansatz:
\begin{equation}\label{ratAnsatz}
    \sum_{\bbeta\in B}p_{\bbeta}(\w)\dd_{\w}^{\bbeta}
  \quad+\quad\tilde{\d}_{v_1}\cdot
    \sum_{\ggamma\in U}\sum_{\aalpha\in A_1}\frac{q_{1,\aalpha,\ggamma}(\w)\,\bm{v}^{\aalpha}}{d_{1,\ggamma}(\bm{v},\w)}
    (\dd_{\bm{v}},\dd_{\w})^{\ggamma} 
  \quad+\quad\dots
\end{equation}
where the notation is as in~\eqref{polyAnsatz} except that
we can restrict the support of the delta part to the monomials under 
the stairs of~$G$ (as in Chyzak's algorithm). The denominators~$d_{i,\ggamma}$
are polynomials in $\K[\bm{v},\w]$. After coefficient comparison with respect 
to~$\bm{v}$ we finally have to solve a linear system
in the $p_{\bbeta}$ and the $q_{i,\aalpha,\ggamma}$ over~$\K(\w)$. This
means that when we will be talking about the denominators $d_{i,\ggamma}$,
we will refer solely to those parts (factors) of the $d_{i,\ggamma}$ that actually
involve some of the variables~$\bm{v}$; the remaining factors will be 
contributed from the solutions~$q_{i,\aalpha,\ggamma}$ over~$\K(\w)$.

Now it is no secret that the denominators $d_{i,\ggamma}$ can be
somehow predicted: We do not know them a priori, but we can deduce a
list of candidate factors that might appear within them. In the
hypergeometric case, Wilf and Zeilberger~\cite{WilfZeilberger92}
already have described how to get a good guess on the denominators. In
general it is not difficult to see that the leading coefficients of
the Gr\"obner basis~$G$ play the key r\^{o}le: consider a solution of
the form~\eqref{polyAnsatz} (that is guaranteed to exist by the
elimination property, provided that~$f$ is holonomic) and reduce its
delta part with~$G$ to obtain a representation of the
form~\eqref{ratAnsatz}.  It is now clear that a solution must exist
where only factors from the leading coefficients of~$G$ as well as
their shifted instances appear in the denominators.

The rest of this section will be dedicated to explaining how this
technique can be completely automatized and made reasonably fast. For
example, the implementations of the hypergeometric case that use
ansatz~\eqref{ratAnsatz} (by
Zeilberger\footnote{http://www.math.rutgers.edu/\~{}zeilberg/programs.html}
in Maple, and by
Wegschaider\footnote{http://www.risc.uni-linz.ac.at/research/combinat/software/MultiSum/}
in\label{refSoftware} Mathematica) require the denominators explicitely
as input. Moreover, Wegschaider in his thesis~\cite{Wegschaider97}
states that ansatz~\eqref{polyAnsatz} is preferable
to~\eqref{ratAnsatz} even when the denominators are known. We do not
share his opinion, as we shall demonstrate below. Or more concretely,
we do not think that this statement holds in the general (not
necessarily hypergeometric) case.

In all ansatz-based algorithms (and hence in our approach), the main
loop is over the support of the principal part. In the following we
concentrate on one single step, i.e., we assume that the set~$B$ is
fixed, and describe our implementation by explaining the various
optimizations that we have included.

\paragraph{Optimization 1} 
To get started with, we need some heuristics for guessing the
denominators~$d_{i,\ggamma}$.  It seems to be natural to start with a
common denominator~$d$, i.e., $d_{i,\ggamma}\mid d$ for all~$i$ and
all~$\ggamma$.  Ideally, a good heuristic should always deliver a~$d$
that is indeed a multiple of all the denominators, but without
overshooting too much.  Experiments suggest that it suffices to take
the denominators that occur during the reduction of the whole ansatz;
recall that its support is already fixed which allows for a very fast
``simulated reduction'' where we do not compute with coefficients but
only with supports.

\paragraph{Optimization 2}
Once we have a good candidate for the common denominator we have to test
whether for this setting, i.e., for the principal part under consideration,
there exists a solution. Additionally we still have to fix the degrees of the
ansatz (the sets~$A_i$); see the next step for this issue. 
To see whether there is a solution we have to reduce the ansatz with
the Gr\"obner basis~$G$ and solve the corresponding linear system. And of
course, it suffices to perform these steps in a homomorphic image.
Hence we plug in some concrete integers for the parameters~$\w$ and reduce
all integer coefficients modulo a 
prime\footnote{in our implementation, we choose 7-digit integers for the~$\w$
and as modulus the largest prime that fits into a machine word, i.e.,
$2147483629$.}. If there is no solution in the 
homomorphic setting, there is, a fortiori, no solution in the general setting.
By choosing these values sufficiently generically we can also minimize the 
risk of obtaining a homomorphic solution that does not extend to a general 
solution. There is one important point to mention and this concerns the 
reduction modulo~$G$: we are working in a noncommutative algebra and therefore
it is problematic to replace an indeterminate by an integer, since by doing so,
we loose the noncommutativity between this indeterminate and the operator that
is connected to it. In~\cite{Koutschan10a} we have described a modular reduction
procedure that keeps track of this issue. The basic idea is that in each step
of the reduction process we have to do the noncommutative multiplication of a
Gr\"obner basis element that makes the leading power products match, in the
general setting, but everything else in the homomorphic setting since no 
noncommutativity is involved any more.

\paragraph{Optimization 3}
We still do not know which $\bm{v}$-degrees in the numerators of our ansatz
we should try. It is manifest to start with small degrees and increase them
until either the homomorphic computations indicate that a solution might exist
or the degrees become unreasonably large (hence here is a second heuristic
involved). The following observation suggests that we can be quite generous with
setting an upper bound for the degrees. Let~$T$ be the ansatz~\eqref{ratAnsatz}
and $T'$ its counterpart with the increased degrees (for this
reasoning it is irrelevant whether we talk about the total degree in~$\bm{v}$
or the componentwise degrees; in any case we have some $A'_i\supseteq A_i,i=1,2,\dots$). 
Then the unknowns that appear in $T'-T$ are precisely the $q_{i,\aalpha_i,\ggamma}$
with $\aalpha_i\in A'_i\setminus A_i$.
Obviously the sets of undetermined coefficients
occurring in $T$ and $T'-T$ have empty intersection. Hence we do not have to 
build the whole linear system from scratch, but instead in each step of the 
degree-looping we have to reduce only the new part $T'-T$ and add some columns 
(and possibly rows) to the matrix. In our implementation we have decided to
loop over an integer~$\delta$ such that $\delta$ limits the degree of each 
of the variables~$\bm{v}$. Once a homomorphic solution is found for a 
certain~$\delta$ we can refine the degree setting componentwise by a few
more modular nullspace computations.

\paragraph{Optimization 4}
Ok, let's now assume that we found a homomorphic solution for some
principal part, some common denominator~$d$ and some degree
setting~$A_i$. We could now use this ansatz to start the final
computation, but there is still lots of possibilities for
improvement. Recall that at this point we only have some common
multiple of the denominators, but not necessarily the minimal
one. Again using homomorphic computations, it is not difficult and not
very costly to figure out the least common multiple of all the
denominators: for each factor of~$d$ we check whether deleting this
factor and decreasing the degree setting of the numerators
accordingly, still yields a solution. If so, we can remove this factor
from our ansatz. Note that such unnecessary factors in the denominator
blow up the degrees of the numerator, too, since they have to cancel
in the end.

\paragraph{Optimization 5}
But should we stop after minimizing the common denominator?  In the
very same fashion, we can now proceed to minimize each single
denominator $d_{i,\ggamma}$. Note that for small examples this
overhead can consume a considerable part of the total computation
time. However, we are definitely convinced that it pays off in big
examples.

\paragraph{Optimization 6}
Last but not least we omit all undetermined coefficients from our ansatz that 
are zero in the homomorphic image. They most probably will be zero in the 
end---but this is folklore\dots

\section{Examples}

In this section we want to present some examples that illustrate the
applicability of our ansatz. To have a fair comparison of the timings,
we do all computations in the same computer algebra system
(Mathematica) and we want to mention that all the code has been
implemented by the same person (unless stated otherwise).
The results are listed in Table~\ref{timings}.
\begin{table}[ht]
\begin{center}
\begin{tabular}{|l||r|r|r||r|r|r|}\hline
              & \multicolumn{3}{|c||}{Ansatz (2.1)} & \multicolumn{3}{|c|}{Ansatz (2.3)} \\ \hline
Example       & time & memory & output & time & memory & output \\ \hline\hline
Bessel        & 127  & 78     & 7.2    & 10   & 1.9    & 7.2    \\ \hline
Gegenbauer    & 1122 & 601    & 13     & 1.6  & 0.66   & 13     \\ \hline
Andrews-Paule & 27   & 30     & 341    & 2.1  & 0.35   & 4.3    \\ \hline
Feynman (wz)  & 81   & 156    & 293    & \multirow{2}{*}{174} & \multirow{2}{*}{85} & \multirow{2}{*}{156} \\
Feynman (zw)  & 171  & 126    & 91     &      &        &        \\ \hline
\end{tabular}
\end{center}
\medskip
\caption{Timings (in seconds), memory usage (in MegaBytes), and output size (in KiloBytes) 
for Chyzak's algorithm and our ansatz; the Feynman example consists of two rows
corresponding to the different orders of integration (in the double sum, changing the summation
order does not lead to different values).}\label{timings}
\end{table}

\subsection{Integral with four Bessel functions}
An example that is often used for testing creative telescoping procedures is the 
following integral over a product of four Bessel functions
\begin{equation}
  \int_0^\infty x J_1(ax) I_1(ax) Y_0(x) K_0(x) \rmd x=-\frac{\log(1-a^4)}{2\pi a^2}.
\end{equation}
The intriguing fact with this example is that the input, the annihilating ideal for the integrand
\[
  \begin{array}{l}
  \{a^3D_{\!a}^4+4a^2D_{\!a}^3-3aD_{\!a}^2+3D_{\!a}+4a^3x^4,\\
  \phantom{\{}x^4D_{\!x}^4-4ax^3D_{\!x}^3D_{\!a}+6a^2x^2D_{\!x}^2D_{\!a}^2
    -4a^3xD_{\!x}D_{\!a}^3+12ax^2D_{\!x}^2D_{\!a}-24a^2xD_{\!x}D_{\!a}^2+\\
  \qquad 8a^3D_{\!a}^3+x^2D_{\!x}^2-26axD_{\!x}D_{\!a}+40a^2D_{\!a}^2-3xD_{\!x}+26aD_{\!a}-4a^4x^4+4x^4+3\}
  \end{array}
\]
as well as the output, the creative telescoping operator
\setlength{\arraycolsep}{0pt}
\begin{equation}\label{besselCT}
  \begin{array}{ll}
  aD_{\!a}+2\quad +& \\
  \quad \displaystyle \D{x}\cdot\frac{1}{4(a^4-1)x^3}\cdot\big(&
    -ax^3D_{\!x}^3D_{\!a}+4a^2x^2D_{\!x}^2D_{\!a}^2-6a^3xD_{\!x}D_{\!a}^3-2x^3D_{\!x}^3\\
    & +12ax^2D_{\!x}^2D_{\!a}-32a^2xD_{\!x}D_{\!a}^2+16a^3D_{\!a}^3-25axD_{\!x}D_{\!a}\\
    & +70a^2D_{\!a}^2-2xD_{\!x}+19aD_{\!a}-16a^4x^4+2\big)
  \end{array}
\end{equation}
are pretty small. In particular, observe that the principal part is of order 1, and 
hence no longish looping is necessary. But the integrand, being a product of four
Bessel functions (which are $\d$-finite with dimension 2), has an annihilating ideal
that contains $16=2\cdot 2\cdot 2\cdot 2$ monomials under its stairs, and this
causes Chyzak's algorithm to take quite long with finding \eqref{besselCT}
(recall that a $16$ by $16$ system of differential equations has to be uncoupled
which causes intermediate expression swell).
Therefore the timings and memory usages differ by more than one order of magnitude:
10s vs. 127s, and 1.9MB vs. 78MB, respectively.

\subsection{A product of three Gegenbauer polynomials}

The following identity can be found as formula (6.8.10) in the book by Andrews, Askey, and Roy~\cite{AndrewsAskeyRoy99}.
It is valid when $\lambda>-\frac{1}{2}$ and $\lambda\neq0$, $l+m+n$ is even and the sum of any two of 
$l,m,n$ is not less than the third. The integral is zero in all other cases:
\[
  \begin{array}{l}
  \displaystyle
  \int_{-1}^1 C_l^{(\lambda)}(x)\, C_m^{(\lambda)}(x)\, C_n^{(\lambda)}(x) \left(1-x^2\right)^{\lambda-1/2} \rmd x=
  \frac{\pi\, 2^{1-2\lambda}\Gamma \left(2\lambda +\frac{1}{2} (l+m+n)\right)}
       {\Gamma (\lambda )^2 \left(\frac{1}{2} (l+m+n)+\lambda \right)}\\
  \displaystyle\qquad\times\frac{(\lambda)_{(m+n-l)/2} (\lambda)_{(l+n-m)/2} (\lambda )_{(l+m-n)/2} }
  {\left(\frac{1}{2} (m+n-l)\right)! \left(\frac{1}{2} (l+n-m)\right)!
   \left(\frac{1}{2} (l+m-n)\right)! (\lambda )_{(l+m+n)/2}}
  \end{array}
\]
One creative telescoping operator that can be used to prove this identity
has the innocent-looking principal part
\[
  (l+m-n+1)(l+2\lambda-m+n-1)S_{\!m}-(l-m+n+1)(l+2\lambda+m-n-1)S_{\!n}.
\]
We fix the support~$\{\S{m},\S{n}\}$ and compare the runtime of
the different ans\"atze.
With our implementation of Chyzak's algorithm we need 1122 seconds to get
the above, whereas our new approach can do it in 1.6 seconds! Again the
reason is that the vector space under the stairs of the Gr\"obner basis
has dimension~$8$ which is already relatively large and leads to huge
intermediate expressions (see Table~\ref{timings}).

\subsection{A hypergeometric double sum}\label{secAndrewsPaule}

The following double sum has been studied by Andrews and Paule~\cite{AndrewsPaule93}
from a human and a computer algebra point of view:
\[
  \sum_i\sum_j\binom{i+j}{i}^2\binom{4n-2i-2j}{2n-2i}=(2n+1)\binom{2n}{n}^2.
\]
This sum being hypergeometric can be treated with both Zeilberger's
Multi-WZ implementation in Maple and Wegschaider's \texttt{MultiSum}
package in Mathematica (see the footnotes on
page~\pageref{refSoftware}). We take the latter
for a comparison. As already mentioned above, we have to give the correct denominators
of the delta parts as input. Then the command \texttt{FindRationalCertificate} takes
0.5s to find the creative telescoping operator
\[
  \begin{array}{ll}1\> -\> &\displaystyle(\S{i}-1)\cdot\frac{i \left(2 i j-i n+i+2 j^2-3 j n+2 j-3 n\right)}{(j+1) (i+j-2 n)}-\\
  &\displaystyle(\S{j}-1)\cdot\frac{j \left(2 i^2+2 i j-3 i n+2 i-j n+j-3 n\right)}{(i+1) (i+j-2 n)}.
  \end{array}
\]
Our implementation, having to figure out the denominators on its own, takes a little
bit longer, namely 2.1s. We believe that this is still reasonably fast and bet that
every user would need more than 1.6 seconds even for only typing the denominators! Using 
Wegschaider's implementation of the ansatz~\eqref{polyAnsatz} (his favourite method)
takes 3.2s and the output is not as nice as the one given above (memory usage: 9MB, 
output size: 200KB).

We have mentioned that Chyzak's algorithm in principle can be iteratively applied to
solve multi-summation problems. This example spectacularly demonstrates that this
strategy can end up in a long, stony way: The creative telescoping relations for
the inner sum are huge and the tricky boundary conditions make things even more difficult.

\subsection{Feynman integrals}

Another interesting area of application is the computation of Feynman integrals
that is a hot topic in particle physics. We borrow a relatively simple example
from the thesis~\cite[(J.17)]{Klein06}:
\[
  \int_0^1\int_0^1\frac{w^{-1-\varepsilon /2}(1-z)^{\varepsilon /2}z^{-\varepsilon /2}}{(z+w-wz)^{1-\varepsilon}}
  \left(1-w^{n+1}-(1-w)^{n+1}\right)\rmd w\rmd z.
\]
The integrand is not hyperexponential and hence the multivariate
Almkvist-Zeil\-berger algorithm is not applicable (at least not
without reformulating the problem). Our implementation needs 174s to
find the third-order recurrence in~$n$ (but can be reduced to 24s by
use of options).  Again we can try Chyzak's algorithm iteratively on
this example, running into similar troubles as in the previous
example, although the swell of the intermediate expressions is by far
not as bad as before (note that the input Gr\"obner basis has only
$3$~monomials under the stairs).  Depending on the order of
integration this takes 81s or 171s, but on the cost of higher memory
consumption (see Table~\ref{timings}).

\section{Conclusion and Outlook}\label{secConclusion}
We have described an approach to finding creative telescoping
relations that is particularly interesting for multiple sums and
integrals as well as for big inputs. For small examples the necessary
preprocessing might well consume most of the computation time, but the
bigger the input is the less this carries weight.  By avoiding the
expensive uncoupling step, our ansatz becomes more attractive as the
size of the set~$U$ (monomials under the stairs) grows.  From a
theoretical point of view Chyzak's algorithm is still preferable since
it is guaranteed to find the smallest creative telescoping relations
(with respect to the support of the principal part) whereas our
approach involves some heuristics which in unlucky cases can prevent
us from getting the minimal output. Therefore we have incorporated
several options into our implementation (e.g., for fixing the support
of the principal part, the common denominator in the delta part or its
numerator degrees) that allow the user to override the built-in
heuristics. In the rare cases where this leads to different results,
we can compute their union (corresponding to the right gcd in the
univariate case) and with high probability end up with the minimal 
telescoper.

We want to conclude with a really big example to which we plan to
apply our method in the near future.  In~\cite{Koutschan10a} we have
presented a computer proof of Stembridge's TSPP theorem using the
polynomial ansatz~\eqref{polyAnsatz}. The computations for achieving
this took several weeks! Attacking the notorious $q$-TSPP conjecture,
which is the $q$-analogue of Stembridge's theorem, therefore seemed to
be hopeless due to the additional indeterminate~$q$ that blows up all
computations by some orders of magnitude. With our fast implementation
of ansatz~\eqref{ratAnsatz} we can now find the creative telescoping
relations for the ordinary TSPP in a few hours! This fact, together
with the previous work done in~\cite{KauersKoutschanZeilberger09b}
makes it likely that the $q$-TSPP conjecture can be turned into a
theorem after being open for about 25 years.

\paragraph{Addendum to the final version} Meanwhile we have succeeded
in proving the long-standing $q$-TSPP conjecture using exactly
the methods described in this paper. The corresponding article
has already been submitted for publication~\cite{KoutschanKauersZeilberger10}.



\subsection*{Acknowledgment}
Some of the ideas described here evolved from inspiring discussions
that I had with Fr\'{e}d\'{e}ric Chyzak and Manuel Kauers during a
visit at INRIA-Rocquencourt. I would like to thank the two anonymous
referees who provided very detailed and helpful reports which
contributed to improve and clarify the exposition significantly.

\end{document}